\documentclass[floats,floatfix,showpacs,amssymb,prd,superscriptaddress,nofootinbib,twocolumn,aps]{revtex4-1}
\usepackage{graphicx, epsfig, amssymb} 
\usepackage{amsmath, amsfonts}
\usepackage{bm} 

\usepackage[linktocpage]{hyperref}
\usepackage[caption=false]{subfig}
\usepackage[usenames]{color}

\def\be{\begin{equation}}
\def\ee{\end{equation}}
\def\beq{\begin{eqnarray}}
\def\eeq{\end{eqnarray}}

\newcommand{\bea}{\begin{eqnarray}}
\newcommand{\eea}{\end{eqnarray}}
\newcommand{\ben}{\begin{enumerate}}
\newcommand{\een}{\end{enumerate}}
\newcommand{\bi}{\begin{itemize}}
\newcommand{\ei}{\end{itemize}}

\def\be{\begin{equation}}
\def\ee{\end{equation}}

\begin{document}

\title{Dynamics of black holes in de Sitter spacetimes}

 \author{Miguel~Zilh\~ao}\email{mzilhao@fc.up.pt}
   \affiliation{Centro de F\'\i sica do Porto, %
   Departamento de F\'\i sica e Astronomia, %
   Faculdade de Ci\^encias da Universidade do Porto, %
   Rua do Campo Alegre, 4169-007 Porto, Portugal%
 }

 \author{Vitor~Cardoso}
   \affiliation{
   Centro Multidisciplinar de Astrof\'\i sica --- CENTRA,
   Departamento de F\'\i sica, Instituto Superior T\'ecnico --- IST,
   Av. Rovisco Pais 1, 1049-001 Lisboa, Portugal
 }
 \affiliation{
   Department of Physics and Astronomy, The University of Mississippi, %
   University, MS 38677-1848, USA
 }

 \author{Leonardo~Gualtieri}
   \affiliation{
     Dipartimento di Fisica, ``Sapienza'' Universit\`a di Roma \& Sezione INFN Roma1, Piazzale Aldo Moro 5, 00185, Roma, Italy}
 
\author{Carlos~Herdeiro}
   \affiliation{
   Departamento de F\'\i sica da Universidade de Aveiro, 
   Campus de Santiago, 3810-183 Aveiro, Portugal
 }

 \author{Ulrich~Sperhake}
   \affiliation{
   Institut de Ci\`encies de l'Espai (CSIC-IEEC), Facultat de Ci\`encies, 
   Campus UAB, E-08193 Bellaterra, Spain
 }
   \affiliation{
   California Institute of Technology,
   Pasadena, CA 91125, USA
 }
   \affiliation{
   Centro Multidisciplinar de Astrof\'\i sica --- CENTRA,
   Departamento de F\'\i sica, Instituto Superior T\'ecnico --- IST,
   Av. Rovisco Pais 1, 1049-001 Lisboa, Portugal  
 }

 \author{Helvi~Witek}
   \affiliation{
   Centro Multidisciplinar de Astrof\'\i sica --- CENTRA,
   Departamento de F\'\i sica, Instituto Superior T\'ecnico --- IST,
   Av. Rovisco Pais 1, 1049-001 Lisboa, Portugal  
 }


\begin{abstract}
Nonlinear dynamics in cosmological backgrounds has the potential to teach us immensely about our universe, and also to serve 
as prototype for nonlinear processes in generic curved spacetimes. Here we report on dynamical evolutions of black holes in asymptotically de Sitter spacetimes. We focus on the head-on collision of equal mass binaries and for the first time compare analytical and perturbative methods with full blown nonlinear simulations. Our results include an accurate determination of the merger/scatter transition (consequence of an expanding background) for small mass binaries and a test of the Cosmic Censorship conjecture, for large mass binaries. We observe that, even starting from small separations, black holes in large mass binaries eventually lose causal contact, in agreement with the conjecture.
 
\end{abstract}


\maketitle
\date{today}
\section{Introduction}
\textit{de Sitter} spacetime is the paradigmatic and, in many ways, the simplest accelerating universe. It is a maximally symmetric solution of Einstein's equations with a positive cosmological constant, describing a Friedmann-Robertson-Walker (FRW) cosmology with a constant Hubble parameter. Moreover,  the large scale structure of our universe appears to be that of a de Sitter geometry, since the large body of observational evidence for a present cosmological acceleration is well modeled by a positive cosmological constant $\Lambda$ \cite{Komatsu:2010fb}. 

Key questions concerning the evolution towards a de Sitter, spatially homogeneous universe are how inhomogeneities develop in time and, in particular, if they are washed away by the cosmological expansion  \cite{Shibata:1993fx}. Answering them, requires controlling the imprint of the gravitational interaction between localized objects on the large-scale expansion. Conversely, the cosmological dynamics should leave imprints in strong gravitational phenomena like primordial black hole (BH) formation \cite{Shibata:1999zs} or the gravitational radiation emitted in a BH binary coalescence, which carry signatures of the cosmological acceleration as it travels across the universe. Identifying these signatures is not only of conceptual interest but also phenomenologically relevant, in view of the ongoing efforts to directly detect gravitational radiation. 

Finally, dynamics in asymptotically de Sitter spacetimes could also teach us about more fundamental questions such as Cosmic Censorship: two BHs of sufficiently large mass in de Sitter spacetime would, upon merger, give rise to too large a BH to fit in its cosmological horizon.
In this case the end state would be a naked singularity. This possibility begs for a time evolution of such a configuration. Does the time evolution
of non-singular data containing two BHs result in a naked singularity, or are potentially offending BHs simply driven away from each other by the cosmological expansion?

Tackling all these issues starts with the study of the coalescence process in BH binaries. Since 2005 \cite{Pretorius:2005gq, Campanelli:2005dd,Baker:2005vv}, extraordinary progress has been achieved in the understanding of BH mergers and the associated gravitational radiation emission, using numerical methods, while the last few years have witnessed a generalized interest in the program of extending numerical evolution of BH spacetimes to generic backgrounds \cite{Cardoso:2012qm,Yoshino:2011zz,Yoshino:2011zza,Sperhake:2011xk,Yoshino:2009xp,Zilhao:2010sr,Witek:2010qc,Bizon:2011gg,Bantilan:2012vu}.
In this paper we shall take the first step to bring these techniques to a new frontier: we report the first numerical evolution of BH binaries in an asymptotically de Sitter geometry. Even though we consider a range of values for the cosmological constant 
far larger than those which are phenomenologically viable, these results provide useful insight on the general features of dynamical BH processes in spacetimes with a cosmological constant, which can improve our understanding of our universe.

\section{Schwarzschild-de Sitter}
\label{sec:mcvittie}
The Schwarzschild-de Sitter spacetime, written in static coordinates reads:
\begin{equation}
ds^2=-f(R)dT^2+f(R)^{-1}dR^2+R^2d\Omega_2 \ . \end{equation}
The solution is characterized by two parameters; the BH mass $m$ and the Hubble parameter $H$ and
\begin{equation}
f(R)=1-2m/R-H^2R^2 \ , \qquad H \equiv \sqrt{\Lambda/3} \ . 
\end{equation}
$f(R)$ has two zeros, at $R=R_\pm$, $R_-<R_+$, if 
\begin{equation} 
0<mH<mH_{\rm crit} \ , \qquad m H_{\rm crit}\equiv \sqrt{1/27} \ . 
\end{equation}
These zeros are the location of the BH event horizon ($R_-$) and of a cosmological horizon ($R_+$). If $H=0$, then $R_-=2m$; if $m=0$, then $R_+=1/H$. If $H,m\neq 0$, then  $R_->2m$ and $R_+<1/H$. Since $R$ is the areal radius, the area of the spatial sections of the cosmological horizon decreases in the presence of a BH; and the area of the spatial sections of the BH horizon increases in the presence of a cosmological constant, as one would intuitively anticipate.

The basic dynamics in this spacetime may be inferred by looking at radial time-like geodesics. They obey the equations $\left(dR/d\tau\right)^2=E^2-f(R)$,  where $\tau$ is the proper time and $E$ is the conserved quantity associated to the Killing vector field $\partial/\partial T$. In the static patch ($R_-<R<R_+$), $E$ can be regarded as energy. From this equation we see that $f(R)$ is an effective potential. This potential has a maximum at 
\be
R_{\rm max}=(m/H^2)^{1/3} \label{eq:rcrit_geo}\,.
\ee
Geodesics starting from rest (\emph{i.e.} $dR/d\tau(\tau=\tau_0)=0$) will fall into the BH if $R_-<R<R_{\rm max}$ or move away from the BH if $R_{\rm max}<R<R_+$.

As we will discuss in the next section, the initial data for an evolution in the de Sitter  universe can be computed 
in a similar manner as has been done 
in asymptotically flat space as long as one chooses a foliation with extrinsic curvature $K_{ij}$ having only a trace part. Such a coordinate system is known for Schwarzschild-de Sitter: \textit{McVittie coordinates} \cite{McVittie:1933zz}. These are obtained from static coordinates by the transformation $(T,R)\rightarrow (t,r)$ given by
\begin{equation}
\label{eq:rtoR}
R = (1+\xi)^2 a(t) r \ , \ \ \  T=t+H \int \frac{RdR}{f(R)\sqrt{1-2m/R}} \ ,
\end{equation}
where $a(t)= \exp(H t)$ and $ \xi \equiv \frac{m}{2 a(t) r}$.
One obtains McVittie's form for Schwarzschild-de Sitter:
\begin{equation}
\label{eq:mcvittie}
ds^2 = - \left( \frac{1-\xi}{1+\xi} \right)^2 dt^2 
+ a(t)^2 (1+\xi)^4 (dr^2 + r^2d\Omega_2 ) \ .
\end{equation}
For $t=\mathrm{constant}$, one can show that 
indeed $K^i_j=-H\delta^i_j$.

By setting $m=0$ in McVittie coordinates one recovers a FRW cosmological model with $k=0$ (flat spatial curvature) and an exponentially growing scale factor. 
The cosmological horizon $\mathcal{H}_C$ discussed above, located at $R=1/H$, stands at $r_{\mathcal{H}_C}=1/(He^{Ht})$.
The spatial sections of $\mathcal{H}_C$ seem to be shrinking down in this coordinate system. What happens, in fact, is that the exponentially fast expansion is taking any observer to the outside of $\mathcal{H}_C$. This is a well known phenomenon in studies of inflation and, as we shall see, has important consequences for the numerical evolution.

\section{Numerical Setup}
\label{sec:evol-eq}
The cosmological constant introduces a new term in the Hamiltonian
constraint obtained after the canonical 3+1 decomposition: $
\,^{3}R~-~K{}_{i}{}_{j}~K{}^{i}{}^{j}~+~K^2~=2\Lambda$,
where ${}^3R$ denotes the Ricci scalar associated with
the spatial three metric and $K_{ij}$, $K$ the extrinsic curvature
and its trace; cf.~\cite{Alcubierre:2008}.
In Refs.~\cite{Nakao:1990vq,Nakao:1992zc} it was observed that imposing
a spacetime slicing obeying $K^i_j=-H\delta^i_j$, and a spatial metric
of the form $dl^2 = \psi^4 \tilde \gamma_{ij} dx^i dx^j$, the equations
to be solved in order to obtain initial data are equivalent to those
in vacuum.  In particular, for a system of $N$ BHs momentarily at rest
(with respect to the given spatial coordinate patch), the conformal
factor $\psi$ takes the form
\begin{equation}
\label{eq:psigen}
\psi = 1 + \sum_{i=1}^N \frac{m_i}{2|r-r_{(i)}|} \,.
\end{equation}
There are $N+1$ asymptotically de Sitter regions, as $|r-r_{(i)}|\rightarrow 0,+\infty$;  the total mass for observers in the common asymptotic region ($|r-r_{(i)}|\rightarrow +\infty$) is $\sum_i m_i$~\cite{Nakao:1992zc}.

For the numerical implementation we make use of the generalized
Baumgarte, Shapiro, Shibata and Nakamura (BSSN) formulation,
\emph{e.g.}~\cite{Alcubierre:2008,Sperhake:2006cy}.
For the case of vanishing cosmological constant the definition
of the BSSN
variables $\chi,\tilde{\gamma}_{ij},\tilde{A}_{ij},\tilde{\Gamma}^k$
and their evolution in time is given by Eqs.~(1, A1-A8)
in Ref.~\cite{Sperhake:2006cy}\footnote{We note a missing factor of $\chi$
in the final term of Eq.~(A6) therein.}\,. For our simulations with
$\Lambda \ne 0$ we apply two modifications to this formalism.
(i) The evolution equation for $K$
becomes $\left( \partial_t -  \mathcal{L}_\beta \right) K  =
[\dots] - \alpha \Lambda$, where $ [\dots] $ denotes the
``standard'' right-hand side of Eq.~(A6) in \cite{Sperhake:2006cy}.
(ii) a new variable $\bar \chi =  \exp(2Ht)  \chi$ has been evolved
instead of $\chi$~\cite{Shibata:1993fx}. The reason is that for
BH evolutions it is crucial to impose a floor value on $\chi$,
typically $10^{-4}$ or $10^{-6}$, which is inconsistent with
the natural behaviour of this variable in a de Sitter spacetime:
$\chi^{-1} \sim \exp(2Ht)$. In contrast $\bar \chi \to 1$ when
$r\to\infty$ for all times.
The evolution equation for $\bar \chi$ is given by
\begin{equation}
\label{eq:evolchibar}
\partial_t \bar \chi = 2 \bar \chi (\alpha K - \partial_i \beta^i )/3
+ \beta^i \partial_i \bar \chi + 2H \bar \chi \,,
\end{equation}
and replaces Eq.~(A6) of \cite{Sperhake:2006cy}.
Boundary conditions for all quantities are imposed by looking at the behavior of massless perturbing fields in a pure de Sitter background.
Accordingly, we impose the following asymptotic behavior for all BSSN variables
\begin{equation}
\label{eq:bcs2}
\partial_t f-\partial_t f_0+\frac{1}{a(t)}\partial_{r}f+\frac{f-f_0}{a(t) r}-H\left (f-f_0\right )=0\,.
\end{equation}
We should note that we also performed evolutions using different sets of boundary conditions, to test the independence of the results on boundary conditions imposed in a region with no causal contact with the interaction region. As far as the behaviour and location of the horizons and all quantities discussed in this paper are concerned, no noticeable difference could be found.

Our numerical simulations use the {\sc Lean} code \cite{Sperhake:2006cy} which
is based on the \textsc{Cactus} computational toolkit~\cite{cactus} and the
\textsc{Carpet} mesh refinement package \cite{Schnetter:2003rb,
  carpet}. The calculation of Black hole Apparent Horizons (BAHs) and Cosmological Apparent Horizons (CAHs) is performed with {\sc AHFinderDirect}
\cite{Thornburg:1995cp, Thornburg:2003sf}. We remark that BAHs, found as marginally trapped surfaces, indicate in the Sitter space (with the same legitimacy as in asymptotically flat space) the existence of an event horizon  \cite{Shiromizu:1993mt}. CAHs are surfaces of zero expansion for  \textit{ingoing} null geodesics. In a single BH case, in McVittie coordinates, the BH event horizon and cosmological horizon are indeed foliated by apparent horizons.

The ``expanding'' behaviour of the coordinate system led us to add a new innermost refinement level at periodic time intervals so as to keep the number of points inside the cosmological horizon approximately unchanged.
The necessity for adding extra refinement levels effectively limits our ability to follow the evolution on very long timescales, as the number of time steps to cover a fixed portion of physical time grows exponentially. This feature resembles in many ways the recently reported work by Pretorius and Lehner on the follow-up of the black string instability \cite{Lehner:2010pn}.

\section{Numerical Results}

\begin{figure}[htbp]
\centering
\includegraphics[width=0.45\textwidth,clip=true]{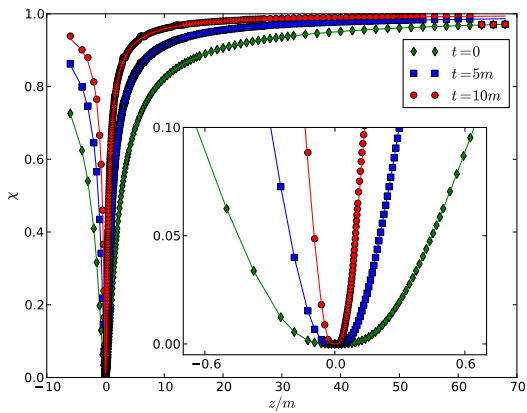}
\caption{
Conformal factor $\chi$ for a single BH evolution with $H=0.8H_{\rm crit}$
using the McVittie slicing condition, Eq. \eqref{eq:mcvittie_slicing}. The
obtained numerical results are plotted, along the $z$ coordinate
(symmetry $\chi(-z)=\chi(z)$ imposed at $z=0$),
against the expected analytical solutions (solid lines). }
\label{fig:mcvittie}
\end{figure}

As a first test on the numerical implementation, we performed evolutions of a single BH imposing the McVittie slicing condition; that is, we use~\eqref{eq:mcvittie} as initial data and impose
\begin{equation}
\label{eq:mcvittie_slicing}
\begin{aligned}
\partial_t \alpha &  = {4 m r H e^{Ht}}/{(m + 2 r e^{Ht})^{2}} \,, \qquad 
\partial_t\beta^i & = 0 \, , 
\end{aligned}
\end{equation}
throughout the evolution. The analytical solution \eqref{eq:mcvittie} can be compared with the numerical results. 
For a single BH evolution with $m=1$ and $H=0.8H_{\rm crit}$, the results are displayed in Fig.~\ref{fig:mcvittie}. 
Using this slicing, the runs eventually crash (at $t \sim 12 m$). By contrast, the standard ``1+log'' slicing condition 
\begin{equation}
\label{eq:1+log}
\partial_t \alpha = \beta^i \partial_i \alpha - 2\alpha \left(
  K - K_0
\right) \,,
\end{equation}
where $K_0 = -3H = -\sqrt{3\Lambda} $, enables us to have long term stable evolutions. As consistency checks, the areal radii at the apparent horizons (both BH horizon and cosmological horizon) are constants in time and have the value expected from the analytical solution in a single BH spacetime. Moreover, the areal radius at fixed coordinate radius evolves with time in the way expected from the exact solution.

\begin{figure}[htbp]
\centerline{\includegraphics[width=0.45\textwidth,clip=true]{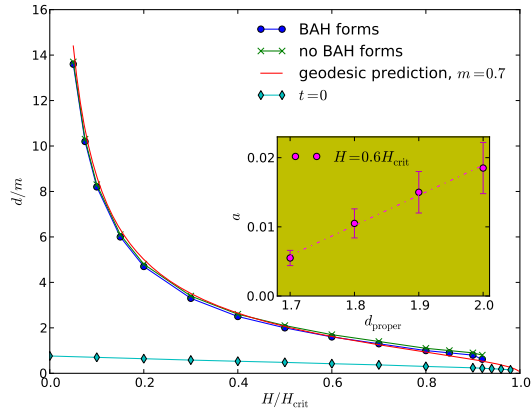}}
\caption[]{\label{fig:crit_sep_P0} Critical coordinate distance for small mass binaries, from both initial data and dynamical evolutions, as well as a point particle estimate, as a function of $H/H_{\rm crit}$. 
We obtain this estimate from the coordinate distance to the horizon, Eq.~\eqref{eq:rcrit_geo},
for a particular value of $m$. The $t=0$ line refers to the critical separation between having or not having a common BAH in the initial data. The inset shows details of the approach to the critical line for $H=0.6 H_{\rm crit}$, where $a$ is an acceleration parameter.}
\end{figure}

For binary BH initial data, we start by reproducing the results of Nakao \emph{et.~al}~\cite{Nakao:1992zc}, where the critical distance between two BHs for the existence of a common BAH already at $t=0$ was studied. We thus prepare initial data \eqref{eq:psigen} with $m_1 = m_2$ and take all quantities in units of the total mass $m = m_1 + m_2$. The two punctures are set initially at symmetric positions along the $z$ axis. The critical value for the cosmological constant, for which the BH and cosmological horizon coincide is now $mH_{\rm crit} = {1}/{\sqrt{27}}$. We call \textit{small (large) mass binaries} those, for which $H<H_{\rm crit}$ ($H>H_{\rm crit}$). Our results for the critical separation in small mass binaries, at $t=0$, as function of the Hubble parameter are shown in Fig.~\ref{fig:crit_sep_P0}. The line (diamond symbols) agrees, after a necessary normalization, with Fig.~14 of~\cite{Nakao:1992zc}. 

\begin{figure}[htbp]
\centering
\includegraphics[width=0.45\textwidth,clip=true]{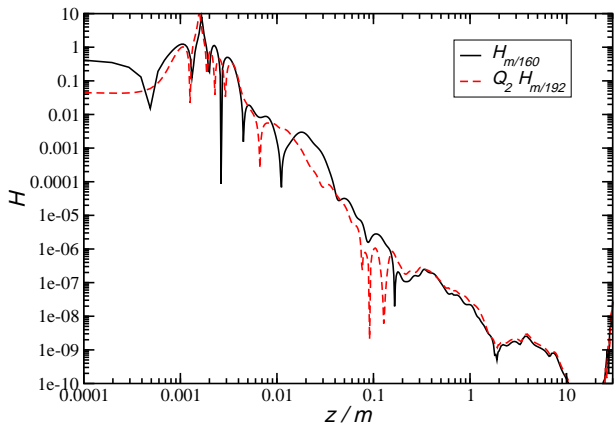}
\caption[]{
Hamiltonian constraint violation along the $z$-axis at time $t=48m$ for
a simulation with $H=0.9H_{\rm crit}$ and initial distance
$d=0.8m$. \label{fig:ham-constraint}}
\end{figure}

We now consider head-on collisions of two BHs with no initial momentum,
\emph{i.e.}~the time evolution of these data. We have monitored the
Hamiltonian constraint violation level for cases with and without
cosmological constant.  We observe that the constraint violations are
comparable in the two cases and plot in Fig.~\ref{fig:ham-constraint}
a snapshot of the Hamiltonian constraint violation at $t=48m$
for parameters $H=0.9H_{\rm crit}$
and $d=0.8m$, a typical case with non-zero
cosmological constant. We have used two resolutions, $m/160$ and
$m/192$ (on the innermost refinement level) and have rescaled the
dashed curve by $Q_2=(192/160)^2$ as expected for second-order
convergence.
%

For subcritical Hubble constant $H<H_{\rm crit} = {1}/{(\sqrt{27}m)}$, we monitor the evolution of the areal radius of the BAHs and that of the CAH of an observer at $z=0$. For instance, for $H=0.9H_{\rm crit}$ and proper (initial) separation $3.69m$ 
we find that the areal radii of the BAH and CAH are approximately constant and equal to $R_{\rm BAH}\simeq 2.36m$ and $R_{\rm CAH}\simeq 4.16m$, respectively. As expected the two initial BAHs, as well as the final horizon, are inside the CAH. As a comparison, a Schwarzschild-de Sitter spacetime with the same $H$ has $R_{\rm BAH}\simeq 2.43m$ and $R_{\rm CAH}\simeq 4.16m$. This suggests that the interaction effects (binding energy and emission of gravitational radiation) are of the order of a few per cent for this configuration.

\begin{figure}[htbp]
\centering
\includegraphics[width=0.45\textwidth]{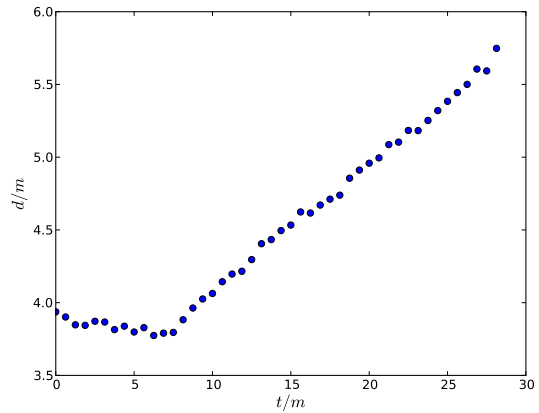}
\caption[]{Proper distance between the BH horizons as a function of
time for the $H=0.9H_{\rm crit}$, and initial (coordinate) distance
$d\simeq 0.9m$. The two holes stay at approximately constant distance
up to $t\approx 8m$ after which cosmological expansion starts
dominating.
\label{fig:proper-dist} }
\end{figure}

As the initial separation grows, so does the total time for merger. For separations larger than a critical value, the two BHs do not merge,
but scatter to infinity. For such scattering configurations, the simulations eventually exhibit a regime of exponentially increasing proper distance between the BAH. 
Just as in scatters of high energy BHs \cite{Sperhake:2009jz}, here
we find that the immediate merger/scatter regimes are separated by a
blurred region, where the holes sit at an almost fixed proper distance
for some time; cf.~Fig.~\ref{fig:proper-dist}.
By performing a large set of simulations for various cosmological parameters
$H$ and initial distance $d$,
we have bracketed
the critical distance for the merger/scatter region as a function of
the Hubble parameter $H$ for the ``dynamical'' case, \emph{i.e.}, the
initial \emph{coordinate} distance between the BHs such that no common
BAH forms.  The results are displayed in  Fig.~\ref{fig:crit_sep_P0}
(circles and $\times$ symbols).

As expected the critical distance becomes larger as compared to the initial data value (``$t=0$'' line): there are configurations for which a common BAH is absent in the initial data but appears during the evolution (just as in asymptotically flat spacetime). The numerical results can be qualitatively well approximated by a point particle prediction - from Eq. \ref{eq:rcrit_geo}. To do such comparison a transformation to McVittie coordinates needs to be done; we have performed such transformation at McVittie time $t=0$. Intriguingly, for a particular value of $m\simeq 0.7$, the point particle approximation matches quantitatively very well the numerical result; the curve obtained from the
geodesic prediction in Fig.~\ref{fig:crit_sep_P0} is barely distinguishable
from the numerical results.

A further interesting feature concerns the approach to the critical line. For an initially static binary close to the critical initial separation, the coordinate distance $d$ scales as $d=d_0+a t^2$. In general the acceleration parameter scales as $\log a=C+\Gamma \log(d-d_0)$,
where $\Gamma=1$ in the geodesic approximation. A fit to our numerical results for $H=0.6 H_{\rm crit}$ (dashed curve in the inset of Fig.~\ref{fig:crit_sep_P0})
for example yields $C=-3.1,\,\Gamma=0.9$ in rough agreement with this expectation. Details of this regime are given in the inset of Fig.~\ref{fig:crit_sep_P0}.

\begin{figure}[htbp]
\centering
\includegraphics[width=0.5\textwidth,clip=true]{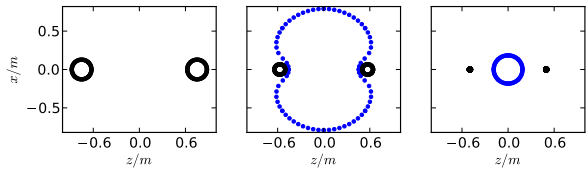}
\caption{Snapshots at different times (from left to right $t/m=0.0, 8.0156, 20.016$) of a simulation with $H= 1.05 H_{\rm crit}$, and an initial coordinate distance $d/m=1.5002$.
The dotted blue line denotes the CAH (for an observer at $z=0$) which is first seen in this simulation at $t/m=8.0156$, highly distorted.
At late times, the CAH has an areal radius of $R = 4.94876$, while the ``theoretical value'' for pure dS is $R=1/H=4.94872$, a remarkable agreement showing spacetime is accurately empty dS for the observer at $z=0$ and the BHs are not in causal contact.
}
\label{horizon_shape}
\end{figure}
Finally, we have performed evolutions with $H~>~H_{\rm crit}$. On the assumption of weak gravitational wave release, such evolutions can test the Cosmic Censorship conjecture since the observation of a merger in such case would reveal a violation of the conjecture~\cite{Hayward:1993tt}. From general arguments and from the simulations with $H<H_{\rm crit}$, we know the cosmological repulsion will dominate for sufficiently large initial distance and in that case we can even expect that a CAH for the observer at $z=0$ will not encompass the BAHs. This indicates the BHs are no longer in causal contact and therefore can never merge. Our numerical results confirm this overall picture. To test the potentially dangerous configurations, we focus on the regime in which the BHs are initially very close. A typical example is depicted in Fig.~\ref{horizon_shape}, for a supercritical cosmological constant $H= 1.05 H_{\rm crit}$, and an initial coordinate distance $d/m=1.5002$. Even though the initial separation is very small, we find that the holes move {\it away} from each other, with a proper separation increasing as the simulation progresses. In fact, further into the evolution, a distorted CAH appears, and remains for as long as the simulation lasts. At late times, this CAH is spherically symmetric, and has an areal radius which agrees, to within $10^{-5}$, with that of an empty de Sitter spacetime with the same cosmological constant.
The evolution therefore indicates that the spacetime becomes, to an excellent approximation, empty de Sitter space for the observer at $z=0$ and that the BHs are not in causal contact. 
Observe that qualitatively similar evolutions can be found in small mass binaries
when the initial distance is larger than the critical value

\section{Final Remarks}

We have presented evidence that the numerical evolution of BH spacetimes in de Sitter universes is under control.
Our results open the door to new studies of strong field gravity in cosmologically interesting scenarios. In closing, we would like to mention that
our results are compatible with Cosmic Censorship in cosmological backgrounds. However, an analytic solution with multiple (charged and extremal) BHs in asymptotically de Sitter spacetime is known, and has been used to study Cosmic Censorship violations \cite{Brill:1993tm}. In \textit{collapsing} universes a potential violation of the conjecture has been reported, although the conclusion relied on singular initial data. To clarify this issue, it would be of great interest to perform numerical evolution of large mass BH binaries, analogous to those performed herein, but in collapsing universes. This will require adaptations of our setup, since the ``expanding'' behaviour discussed of the coordinate system will turn into a ``collapsing'' one, which raises new numerical challenges.


\section*{Acknowledgements}
We would like to thank J.~Thornburg for his help with {\sc AHFinderDirect}.
M.Z.\ would like to thank the hospitality of the Perimeter Institute for Theoretical Physics, 
through the Visiting Graduate Fellows program, where this work was finished. The authors thank the Yukawa Institute for Theoretical Physics at Kyoto University,
 where this work was completed during the YITP-T-11-08 on ``Recent advances in numerical 
and analytical methods for black hole dynamics''.
M.Z.\ and H.W.\ are funded by FCT through grants SFRH/BD/43558/2008 and
  SFRH/BD/46061/2008.  U.S.\ acknowledges support from the Ram\'on y Cajal
  Programme, Grant FIS2011-30145-C03-03
  of the Ministry of Education and Science of Spain
  and Grant AGAUR-2009-SGR-935.
  This work was supported by the {\it DyBHo--256667} ERC Starting Grant,
  the {\it CBHEO--293412} FP7-PEOPLE-2011-CIG Grant,
  the {\it NRHEP--295189} FP7-PEOPLE-2011-IRSES Grant, and by
  FCT -- Portugal through projects PTDC/FIS/098025/2008, PTDC/FIS/098032/2008, PTDC/FIS/116625/2010,  CERN/FP/116341/2010  and the Sherman Fairchild Foundation to Caltech.
  Computations were performed with the support of the
  NSF TeraGrid and XSEDE Grant PHY-090003,
  PRACE-2IP DECI-7 Project ``Black hole dynamics in metric theories of gravity'',
  RES Grants AECT-2012-1-0008, AECT-2011-3-0007, AECT-2011-2-0015 and AECT-2011-3-0006 through the Barcelona Supercomputing Center,
  CESGA Grants ICTS-200 and ICTS-221,
  on the Blafis cluster at Aveiro University, the Milipeia in Coimbra and the Lage cluster at Centro de F\'\i sica do Porto.


\end{document}